\documentclass[twocolumn,showpacs,prb,amsmath,amssymb,floatfix,preprintnumbers]{revtex4}

\usepackage{graphicx}
\usepackage{dcolumn}
\usepackage{bm}

\begin{document}

\title{
Nanoscale structures formed in silicon cleavage
studied with large-scale electronic structure calculations; 
surface reconstruction, step and bending.
}
\author{Takeo Hoshi$^{1,2}$} 
\author{Yusuke Iguchi$^1$}
\author{Takeo Fujiwara$^{1,2}$}
\affiliation{$^1$Department of Applied Physics, University of Tokyo,   
Bunkyo-ku, Tokyo, Japan. \\
$^2$Core Research for Evolutional Science and Technology (CREST),
Japan Science and Technology Agency (JST),
Kawaguchi-shi, Saitama, Japan.
}
\date{\today}

\begin{abstract}

The 10-nm-scale structure 
in silicon cleavage is studied by 
the quantum mechanical calculations 
for large-scale electronic structure. 
The cleavage process on the order of 10 ps shows
surface reconstruction and step formation.
These processes are studied by analyzing electronic freedom 
and compared with STM experiments.
The discussion presents the stability mechanism of 
the experimentally observed mode,
the $(111)$-$(2 \times 1)$ mode,
beyond the traditional approach with surface energy.
Moreover, in several results,
the cleavage path is bent into the experimentally observed planes,
owing to the relative stability among different cleavage modes.
Finally, several common aspects 
between cleavage and other phenomena 
are discussed
from the viewpoints of 
the nonequilibrium process and the 10-nm-scale structure.
\end{abstract} 
\maketitle 


\section{Introduction \label{INTRODUCTION}}

Cleavage is a nonequilibrium process and 
its dynamical mechanism is essential.
In particular, 
the cleavage of silicon single crystals 
is of great interest from the multiscale viewpoint
between macroscale and atomicscale pictures.
In the macroscale picture,
silicon shows perfect brittleness
and brittle fracture is usually
described by continuum mechanics. 
\cite{GRIFFITH,FRAC-MOTT,BRITTLE}
In the atomicscale picture, on the other hand,
a cleaved surface contains areas 
with well-defined reconstructions. 
Currently,
these atomicscale structures are observed by 
scanning tunneling microscopy (STM) \cite{NEDDER}
and other experiments. 
The multiscale feature of the phenomenon,
as discussed in this paper,
appears in 
nanoscale (or 10 nm scale) processes
near the crack tip,
though such processes
cannot be seen by direct ({\it in situ}) experimental observation.

A fundamental question is 
what Miller index and 
surface reconstruction appear at the cleavage surface.
A traditional prediction is 
that the cleavage plane should be
that with the smallest surface energy,
or the smallest energy loss with forming surface.
This prediction, however, is actually not satisfactory 
owing to the following experimental facts;
(i) The easiest cleaved surface of Si is 
a metastable (111)-$(2 \times 1)$ structure
\cite{NEDDER,PANDEY,NORTHRUP,SI111-21-PARRINELLO-MD,
HAUNG,SPENCE93,ROHLFING99}
and will change, irreversibly,
to the ground-state $(7 \times 7)$ structure.
\cite{HENZLER73,NEDDER}
(ii) The (110) cleavage plane 
is also experimentally observed but less favorable
(see an experimental (STM) study, ~\cite{FEENSTRA95-SI110}
and a recent theoretical study \cite{GUMBSCH}).
(iii) 
The cleaved Ge(111) surface also shows the same $(2 \times 1)$
structure,
while the ground state surface structure 
is the c$(2 \times 8)$ structure. 
\cite{NEDDER, STEKOL-2, NOTE-GE-111}
These facts imply the importance of 
direct cleavage simulations with electronic structure.
Although such simulations
have been carried out thus far, \cite{HAUNG,SPENCE93,GUMBSCH} 
the investigation is still limited, 
owing to the system size of $10^2$ atoms.

In this paper, 
the cleavage of silicon is studied with
quantum mechanical calculations
for large-scale electronic structures 
\cite{HOSHI2000A,HOSHI2001A,HOSHI2003A,GESHI,TAKAYAMA2004A,THESIS} 
and  we use a transferable Hamiltonian \cite{KWON}  
in the Slater-Koster (tight-binding) form.
The methodology is reviewed briefly
in Appendix \ref{APPENDIX-A}. 
The method realizes
the cleavage process
with more than $10^4$ atoms
or a sample length of 10 nm.
This paper is organized as follows;
Section \ref{PROBLEM} describes 
the important aspects discussed in this paper.
The easiest cleavage mode on the $(111)$-$(2 \! \times \! 1)$ plane 
is discussed in Section \ref{SI111} and Section \ref{SI111-STEP}.
The latter section focuses on step formations.
The simulation results for bending in the cleavage path
are presented in Section \ref{SECTION-BEND}.
Finally, in Section \ref{DISCUSSIONS},
several common aspects
of nanoscale structures are discussed
for cleavage and other phenomena.


\begin{figure}[thb]
\begin{center}
 \includegraphics[width=7cm]{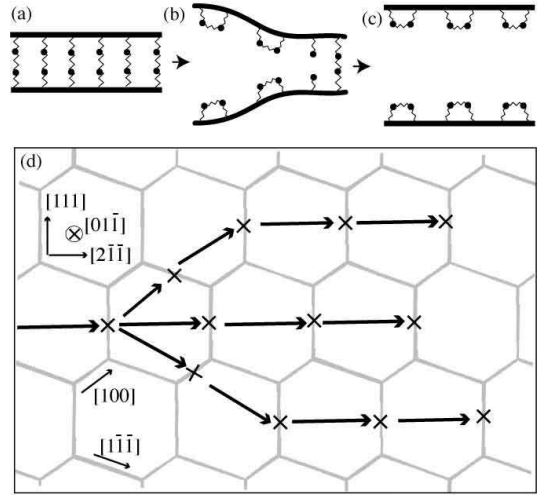}
\end{center}
\caption{
(a)-(c): Toy model of cleavage,
in which surface reconstruction 
is illustrated as dimerization.
(d): Possible cleavage paths on Si(111) plane.
}
\label{FIG-GEOMETRY}
\end{figure}

\begin{figure*}[thb]
\begin{center}
 \includegraphics[width=14cm]{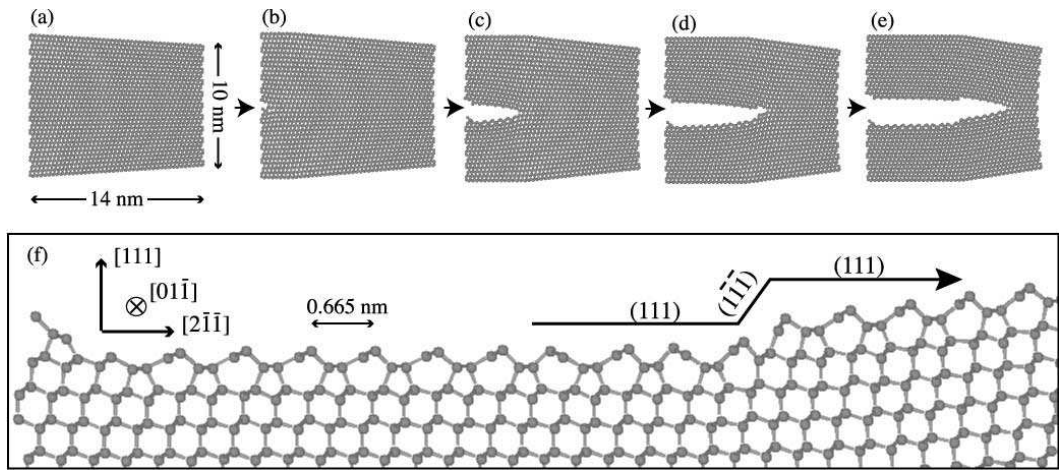}
\end{center}
\caption{
Cleavage process of silicon 
with $(111)$-$(2 \times 1)$ cleaved surface.
(a)-(e):Successive snapshots 
with a time interval of approximately 2 ps.
(f): A part of the lower cleavage surface shown in the snapshot (e).
A step is found and is classified into 
the \lq $[\bar{2}11]$-type' or \lq via-$(1\bar{1}\bar{1})$-plane type'
that can be decomposed into 
the successive bending of cleavage planes as
$(111) \rightarrow (1\bar{1}\bar{1}) \rightarrow (111)$.
}
\label{FIG-111-LARGE}
\end{figure*}


\section{Aspects of cleavage process in real crystal \label{PROBLEM}}

The first aspect of cleavage 
is the typical time scale
as a nonequilibrium process. 
The time scale is 
determined by the cleavage propagation velocity.
In the continuum mechanics and many experiments,
the propagation velocity is given on the order of,
but less than, 
the sound velocity or the Rayleigh wave velocity
($v_{\rm R}= 4.5 {\rm km/s} = 4.5 {\rm nm/ps}$ for Si). 
\cite{FRAC-MOTT,BRITTLE,VELOCITY}
Since the atomistic process occurs within the time scale,
the reconstruction in cleavage, unlike that in annealing, 
should occur {\it locally}. 
In a typical process,
a surface-bound state is formed between electrons in 
{\it nearest-neighbor} dangling bond sites,
as illustrated in Figs.~\ref{FIG-GEOMETRY}(a)-(c).
In other words,
the elementary process should contain 
{\it two} successive bond breakings, {\it not one}.
This nearest-neighbor reconstruction mechanism will 
directly give the experimentally observed 
$(2 \times 1)$ reconstruction (See Section \ref{SI111}).

The second aspect comes from the crystal structure.
Figure ~\ref{FIG-GEOMETRY}(d) shows
possible cleavage paths on the Si$(111)$ plane,
with or without step formation.
Unlike the toy model of Figs.~\ref{FIG-GEOMETRY}(a)-(c),
the system does not have 
the mirror symmetry with the cleavage plane 
and the upper and lower cleaved surfaces
are inequivalent in symmetry.
In particular,
the inequivalence 
between the upper and lower stepped paths
in Fig.~\ref{FIG-GEOMETRY}(d) 
is distinctive in experiments.

Since experiments reported
only the (111) or (110) cleavage plane,
theory should explain
why other surfaces {\it do not} appear.
An interesting fact is that
the calculated surface energy of 
the reconstructed (001) surface 
is {\it smaller} than that of the (111)-$(2 \times 1)$ surface
($\gamma_{(001)}^{4 \times 2} = 1.41 {\rm J/m^2}
 < \gamma_{(111)}^{2 \times 1} = 1.44 {\rm J/m^2}$)
\cite{STEKOL-2,NOTE-SURFACE-ENERGY}.
This means that the absence of the (001) cleavage surface 
is not simply predicted by surface energy.
\cite{NOTE-SURFACE-ENERGY-ORDER} 
The possibility of the (001) cleavage mode
was investigated in our previous study, \cite{HOSHI2003A}
in which the (001) cleavage planes are observed 
for small sample sizes less than 10 nm.
For larger sizes, 
the flat (001) cleavage surface becomes fairly unstable 
and many steps form.
The instability of the $(001)$ cleavage mode is helpful
in understanding the stability of the $(111)$ cleavage mode,
as discussed in Section \ref{SI111-STEP}.

\begin{figure*}[thb]
\begin{center}
  \includegraphics[width=17cm]{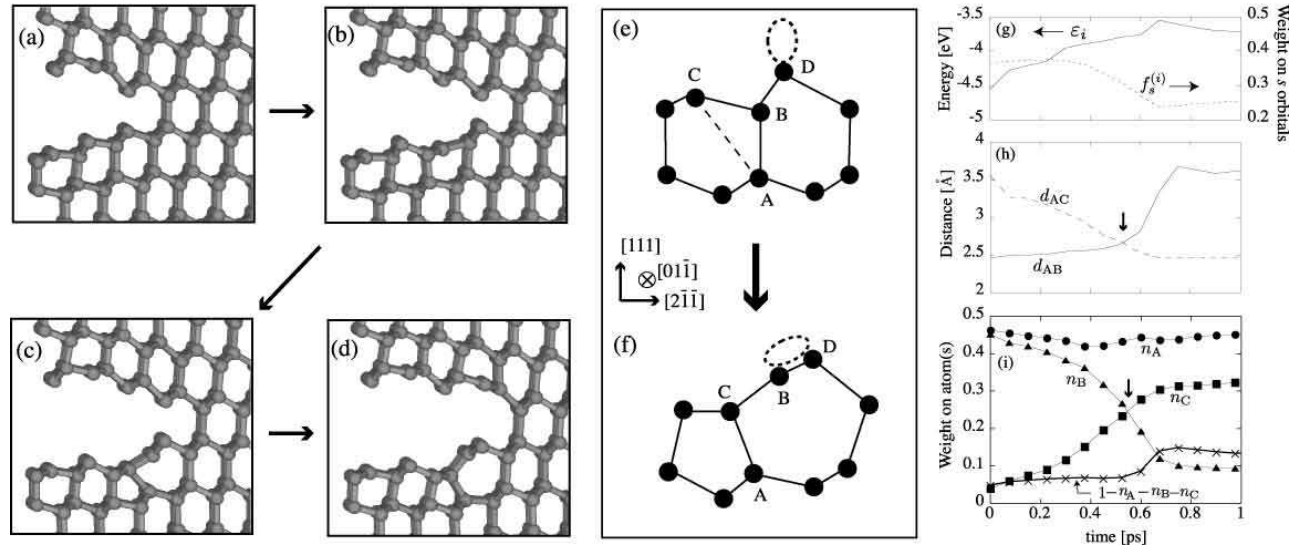}
\end{center}
\caption{
(a)-(d):Initiation of cleavage 
while forming the $(111)$-$(2 \times 1)$ structure. 
(e)-(f): Schematic figures of the transition 
from (e) the buckled $(2 \times 1)$ structure 
to (f) the (tilted) $\pi$-bonded $(2 \times 1)$ structure. 
The former structure appears 
on the lower cleaved surface in (b) and the latter appears in (d).
The oval in (e) indicates the presence of 
a lone pair state on the $D$ atom site.
The oval in (f) indicates 
the $\pi$-bonding zigzag chain 
in the direction perpendicular to the page.
(g)-(i): Quantum mechanical analysis of a process in which 
the bonding wavefunction $(\phi_i)$ between $A$ and $B$ sites 
changes into that between $A$ and $C$ sites;
(g) The one-electron energy 
$\varepsilon_i \equiv \langle \phi_i | H | \phi_i \rangle$
and the weight of $s$ orbitals  $f_s^{(i)}$
for the wavefunction $\phi_i$.
(h) The atomic distance between $A$ and $B$ sites ($d_{AB}$) 
and that between $A$ and $C$ sites ($d_{AC}$). 
(i) The spatial weight distribution 
on the $A$, $B$ and $C$ atom sites 
($n_{\rm A}$, $n_{\rm B}$, $n_{\rm C}$)
for the wavefunction $\phi_i$.
The rest of the weight $(1-n_{\rm A}-n_{\rm B}-n_{\rm C})$
is also plotted.
}
\label{FIG-HANE-PAN-1}
\end{figure*}


\section{$(111)$-$(2 \times 1)$ cleavage mode \label{SI111}}

In experiments,
the cleavage on the Si(111)-$(2 \times 1)$ plane 
is the easiest cleavage mode.
As the atomic structure of the cleaved surface,
Pandey's $\pi$-bonded structure \cite{PANDEY}
is now widely accepted. 
\cite{NEDDER,NORTHRUP,SI111-21-PARRINELLO-MD,HAUNG,
SPENCE93,ROHLFING99}
An actual surface formation process was shown,\cite{HAUNG} 
when a {\it parallel} separation is introduced 
in a slab with the minimal periodic simulation cell
for the $(2 \times 1)$ structure
(See Figs. 1 and 2 of Ref.~\cite{HAUNG}).

The present cleavage simulations are performed
with the $[2\bar{1}\bar{1}]$ propagation directions 
(Fig.~\ref{FIG-GEOMETRY}(d) for geometry),
which are consistent with those in typical experiments
(See Fig. 1 of Ref.~\cite{MERA92}, for example).
An external load is imposed and 
its physical origin 
is the concentrated elastic field 
in the macroscale experimental sample.
See Appendix \ref{DETAIL-CLEAVE} for details.
A smaller sample with $416$ atoms
was also simulated 
as shown in Appendix \ref{TOUGHNESS},
so as to confirm the quantitative agreement
with previous experimental or theoretical studies.
In all cases, 
the periodic boundary condition is imposed 
in the $[01\bar{1}]$ direction,
which is orthogonal to the cleavage propagation direction.
The periodic length contains eight atomic layers 
or is four times larger 
than that of the minimum crystalline periodicity.
In some simulations,
the atomic motion is 
under a constraint 
by the minimum crystalline periodicity in the $[01\bar{1}]$ direction.
We call this constraint, the \lq 2D-like constraint'.
For a systematic investigation,
simulations were carried out 
with and without the \lq 2D-like' constraint.

Figure~\ref{FIG-111-LARGE} shows an example 
of the simulation within the \lq 2D-like' constraint, 
in which a step appears.
The sample contains 11,096 atoms.
Hereafter, 
the bonds (rods) are drawn just as a guide for eye.
From Fig.~\ref{FIG-111-LARGE},
the cleavage propagation velocity is estimated 
to be $v_{\rm prop} \approx 2$nm/ps=2km/s, 
as expected in Section \ref{PROBLEM}.
The cleaved surface in Fig.~\ref{FIG-111-LARGE}(f)
contains the $\pi$-bonded $(111)$-$(2 \times 1)$ structure,
of which unit structure is a set of seven- and five-membered rings.

\begin{figure}[thb]
\begin{center}
  \includegraphics[width=7cm]{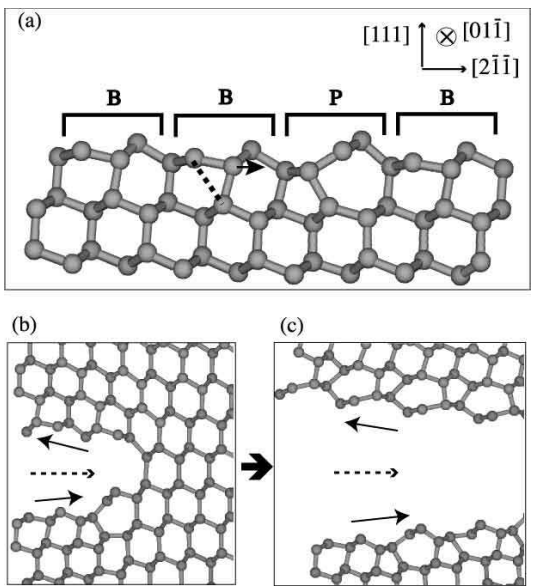}
\end{center}
\caption{
(a) Si(111) surface with coexistence of 
the buckled and $\pi$-bonded $(2 \times 1)$ structures.
The induced strain is shown as an arrow 
at the boundary between
the buckled and $\pi$-bonded structures.
The dashed line indicates the bond site 
that will appear after the next induced BP transition.
(b)-(c): Two snapshots of cleavage process. 
The solid arrow corresponds to 
the direction of the possible surface strain force shown in (a).
The dashed arrow denotes the cleavage propagation direction.
}
\label{FIG-ADD-SHEAR}
\end{figure}
%

\subsection{Elementary reconstruction process \label{SI111-ELEM}}

Figures~\ref{FIG-HANE-PAN-1}(a)-(d) show 
an example of $(2 \times 1)$ reconstruction.
As in the previous simulation with the parallel separation,
\cite{HAUNG}
the reconstruction process has two stages;
The buckled $(2 \times 1)$ structure
of Fig.~\ref{FIG-HANE-PAN-1}(e) appears, 
as shown in Fig.~\ref{FIG-HANE-PAN-1}(b),
when the two dangling-bond electrons at the $C$ and $D$ atom sites
form an atomic (lone-pair) state at the $D$ atom site.~\cite{HANEMAN}  
After that,
the $\pi$-bonded $(2 \times 1)$ structure of Fig.~\ref{FIG-HANE-PAN-1}(f)
appears, since 
(i) the atomic state is transformed into the $\pi$-bonding state
between the $B$ and $D$ sites and 
(ii) the bonding state between the $A$ and $B$ sites 
is transformed into one between the $A$ and $C$ sites,
so as to reproduce the tetrahedral coordination in the $C$ site.
In the three-dimensional view, 
a $\pi$-bonding zigzag chain appears 
in the direction perpendicular to the page.
Hereafter, 
the transition from the buckled structure into the $\pi$-bonded one is called 
the \lq BP transition'.
Note that the resultant $\pi$-bonded structures are always tilted, 
since the $D$ or $B$ atom site is shifted to the vacuum side.
The two tilted structures are inequivalent in symmetry and 
have a small difference in surface energy. 
\cite{STEKOL-2,ROHLFING99} 
In this paper, however,
we do not focus on
the difference between the two tilted structures.

The above explanation of (ii)
is confirmed by the following quantum mechanical analysis;
Figure ~\ref{FIG-HANE-PAN-1}(g) shows
the one-electron energy 
$\varepsilon_i \equiv \langle \phi_i | H | \phi_i \rangle$
and the weight of $s$ orbitals $f_s^{(i)}$.
\cite{HOSHI2003A}
For example, $f_{\rm s}^{(i)} = 1/4$ in an ideal $sp^3$ hybridized state.
Figure ~\ref{FIG-HANE-PAN-1}(h) shows
the distances between sites $A$, $B$ and $C$.
Figure ~\ref{FIG-HANE-PAN-1}(i) shows
the spatial weight distribution 
($|\phi_i(\bm{r})|^2$) on atom sites.
Note that the intermediate wavefunction, 
indicated as the arrow in 
Fig.~\ref{FIG-HANE-PAN-1}(h) or (i),
shows no characteristic feature
in $\varepsilon_i$ and $f_s^{(i)}$,
such as maxima, minima or plateaus.
This is quite different from 
the reconstruction process on the $(001)$ surface,
\cite{HOSHI2003A}
in which the intermediate wavefunction 
shows a plateau with 
a large weight on the $s$ orbitals ($f_{s}^{(i)} \approx 0.8$).


\subsection{Role of anisotropic surface strains \label{SI111-STRAIN}}

We studied the role of 
anisotropic surface strains
in the reconstruction process. 
As a demonstration,
Fig.~\ref{FIG-ADD-SHEAR}(a) 
shows the $(111)$ surface 
with the coexistence of 
the buckled and $\pi$-bonded $(2 \times 1)$ structures.
An anisotropic strain force is induced, which is depicted by an arrow.
The strain force enhances a further BP transition of the next left unit.
Owing to the crystalline symmetry,
the strain force 
on the upper and lower cleaved surfaces
shows the opposite direction to each other,
as shown by solid arrows in Fig.~\ref{FIG-ADD-SHEAR}(b).
This explains that  
the $\pi$-bonded structure appears 
on the lower cleavage surface in Fig.~\ref{FIG-HANE-PAN-1}(d),
but not on the upper one.
Since the experiments on cleaved samples
support the $\pi$-bonded structure, \cite{NORTHRUP,ROHLFING99}
the $\pi$-bonded structure should appear 
commonly on the upper and lower surfaces.
Such a result is obtained in Fig.~\ref{FIG-ADD-SHEAR}(c), 
in which 
an additional constant force is imposed in the left direction
on several leftmost atoms of the upper cleaved surface. 
The additional force is created 
by the boundary condition of surface strain, 
since the simulation sample should be {\it embedded} 
in a macroscale sample.
In the {\it embedded} situation,
the boundary region of the simulation sample 
should be connected with successive $\pi$-bonded $(2 \times 1)$ structures 
and should be under the resultant strain force.
In Fig.~\ref{FIG-ADD-SHEAR}(c),
the resultant sample contains 
the $\pi$-bonded $(2 \times 1)$ structures 
on the upper and lower surfaces.

We also found that 
the sample geometry can have an effect 
on the possible BP transition,
through strain.
For example, 
the BP transition seems to occur more easily
in a {\it thicker} sample,
with a larger sample length in the $[111]$ direction,
since the BP transition should accompany the strain 
in deeper (subsurface) layers.
This observation was unchanged,
when additional simulations were carried out by more complicated Hamiltonians 
with a better reproduction of the surface energy.
A more systematic investigation should be carried out, in the future,
on the quantitative condition of the BP transition.

\section{Step formation in cleavage  \label{SI111-STEP}}


Several (111) cleavage simulations 
contain step formation. 
There are two inequivalent paths
shown as the upper and lower paths in Fig.~\ref{FIG-GEOMETRY}(d). 
The \lq upper' type of path 
is usually called the \lq $[\bar{2}11]$ type',
since the step is {\it descending} in the $[\bar{2}11]$ direction.
In the same manner,
the \lq lower' type is called the \lq $[2\bar{1}\bar{1}]$ type'.
The two types of step are experimentally reported.
In early papers, \cite{HENZLER73,FEENSTRA87,NEDDER} 
it was reported 
that the {\it lower} path in Fig.~\ref{FIG-GEOMETRY}(d), 
the \lq $[2\bar{1}\bar{1}]$ type' is predominant.
In later experimental papers~\cite{TOKUMOTO}, however,
the {\it upper} path, i.e.
the \lq $[\bar{2}11]$' type step, was reported.
In our simulations, 
both types of step are observed.

\subsection{ \lq Upper'-type step \label{SI111-STEP1}}

\begin{figure}[thb]
\begin{center}
 \includegraphics[width=7cm]{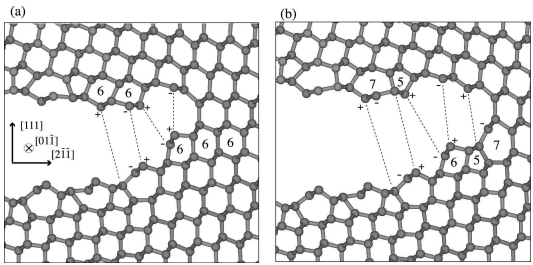}
\end{center}
\caption{
Step formation process on (111) cleavage plane.
The step is classified into 
the \lq $[\bar{2}11]$-' or \lq via-$(1\bar{1}\bar{1})$-plane'-type step 
(See Fig.~\ref{FIG-111-LARGE}(f)).
Two snapshots with a time interval of approximately 0.6 ps are shown.
The dashed lines indicate the initial (crystalline) bonds.
The \lq +' or \lq -' symbols on atoms indicate
the excess and deficit electron populations, respectively.
}
\label{FIG-111-STEP-INPLANE}
\end{figure}

First, 
we discuss the step in Fig.~\ref{FIG-111-LARGE},
which is classified into 
the \lq upper' path 
in Fig.~\ref{FIG-GEOMETRY}(d) or  
the $[\bar{2}11]$ type.
This type of step was observed with and without the \lq 2D-like' constraint.
The formation process is
shown in Fig.~\ref{FIG-111-STEP-INPLANE}.
Hereafter, 
the members of several rings are plotted, 
such as \lq 6', \lq 5'  or \lq 7',
so as to clarify the reconstruction.
This step can be decomposed into 
the successive bendings in the cleavage path
as $(111) \rightarrow (1\bar{1}\bar{1}) \rightarrow (111)$ planes 
(See Fig.~\ref{FIG-GEOMETRY}(d) and Fig.~\ref{FIG-111-LARGE}).
That is,
the step is formed {\it through} the $(1\bar{1}\bar{1})$ plane
and we call the step, the \lq via-$(1\bar{1}\bar{1})$-plane' type.
In the crystalline geometry,
the $(111)$ and $(1\bar{1}\bar{1})$ planes are equivalent. 
If we ignore the reconstruction 
(or quantum mechanical) freedom, 
the step formation can be understood,
since bending between equivalent planes
can occur by 
the local fluctuation of 
the (concentrated) elastic field at the crack tip, 
particularly in its angular dependence.

As a quantum-mechanical analysis,
snapshots of the step formation are shown 
in Fig.~\ref{FIG-111-STEP-INPLANE},
in which an atom with an excess or deficit electron population 
is marked \lq +' or \lq -', respectively.
The initial bonding states (dashed line)
changed into the surface-bound state mainly localized 
on a  \lq +' atom site.
An atom \lq +' 
is always placed at the vacuum side of a buckled structure,
which can be understood by 
the general quantum mechanical tendency. \cite{NOTE-TILT}
The resultant surface shows a balanced structure,
owing to the alternate alignment of the \lq +' and \lq -' sites.
The present step structure is concluded to be a possible one,
although it has not yet been confirmed experimentally.

Now we comment on 
a pioneering theoretical paper,
by Chadi and Chelikowsky, \cite{CHADI-CHELIKOW}
in which the \lq upper' and \lq lower' types of step are compared 
with assumed atomic structures.
They concluded that 
the {\it upper} path is unrealistic,
which seems to be in contrast to the present result.
The contrast appears, because
the present step structure is different 
from that assumed in Ref.~\cite{CHADI-CHELIKOW}
In this reference,
the dangling bonds at the step edge are assumed to be rebonded with each other
in {\it the perpendicular direction} of the page of
Fig.~\ref{FIG-GEOMETRY}(d),
but such a rebonding process results in only a tiny energy gain. 
In the present step structure of  
Fig.~\ref{FIG-111-STEP-INPLANE},
on the other hand,
the dangling bond electrons at the step edge are rebonded 
{\it within the plane} of Fig.~\ref{FIG-GEOMETRY}(d),
whose energy gain mechanism is explained above.

\begin{figure}[tb]
\begin{center}
 \includegraphics[width=7cm]{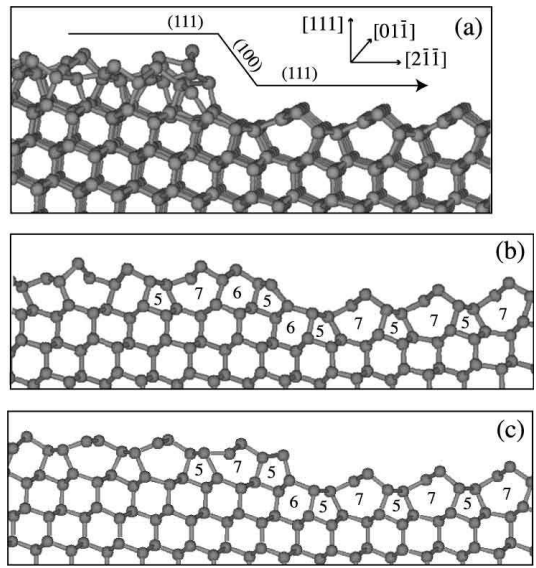}
\end{center}
\caption{
(a) (111) cleaved  surface with
\lq $[2\bar{1}\bar{1}]$'- or 
\lq via-(100)-plane'-type  step.
The arrow indicates the cleavage  propagation direction. 
(b)(c): Two \lq slices' of (a) (See text for details).
}
\label{FIG-STEP-111-2}
\end{figure}

\subsection{\lq Lower'-type step  \label{SI111-STEP2}}

Several simulations 
result in the appearance of 
the other type of step, 
the {\it lower-type} step in Fig.~\ref{FIG-GEOMETRY}(d)
or the \lq $[2\bar{1}\bar{1}]$' type,
when they were carried out 
without the \lq 2D-like' constraint
or minimal periodicity in the $[01\bar{1}]$ direction.
An example is shown in Fig.~\ref{FIG-STEP-111-2}(a).
To observe the freedom in the $[01\bar{1}]$ direction,
we classify the atoms, by their initial positions,
into subsystems or \lq slices'.
Each \lq slice' contains atoms 
within the minimum unit length (two atom layers) 
in the $[01\bar{1}]$ direction.
Four slices are defined
in Fig.~\ref{FIG-STEP-111-2}(a) 
and two of them are shown in Figs.~\ref{FIG-STEP-111-2}(b) and (c).

\begin{figure}[tb]
\begin{center}
  \includegraphics[width=7cm]{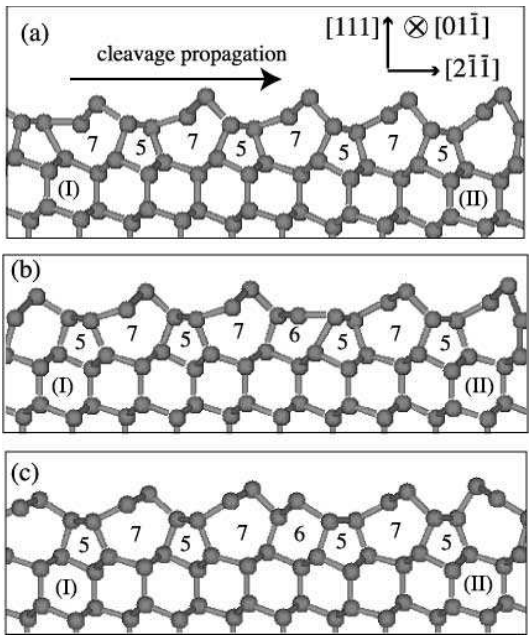}
\end{center}
\caption{
Defective area on cleaved Si(111) surface. 
Three slices are picked out. 
The six-membered rings marked (I) and (II) are geometrically 
equivalent among slices (a)-(c). 
}
\label{FIG-111-DEFECT}
\end{figure}

In Fig.~\ref{FIG-STEP-111-2}(a),
the cleaved surface is defective 
before the step formation. 
In the defective area of Figs.~\ref{FIG-STEP-111-2}(b) and (c),
six-membered rings appear in the surface layer,
as well as five- or seven-membered rings.
A defective area also appears in flat (nonstepped) areas,
as in Fig.~\ref{FIG-111-DEFECT}.
The appearance of the defective six-membered rings means 
that the reconstruction of the dangling-bond electrons 
does not occur  
{\it within the slice}.
We should recall 
that an ideal $(111)$ surface has 
symmetry with a $\pm 2 \pi/3$ rotation and 
the $(2 \times 1)$ reconstruction mechanism shown in
Fig.\ref{FIG-HANE-PAN-1}(e) or (f) is possible in 
the $[2\bar{1}\bar{1}]$, $[\bar{1}2\bar{1}]$ and 
$[\bar{1}\bar{1}2]$ directions.

This type of step formation is decomposed into 
the successive bendings of cleavage
as $(111) \rightarrow (100) \rightarrow (111)$ planes.
In other words,
the step is formed {\it through} the $(100)$ plane
and we call the step the \lq via-$(100)$-plane' type
(See Fig.~\ref{FIG-GEOMETRY}(d) for geometry).
At the step edge,
the dangling bond sites are equivalent to those 
on a (100) or (001) plane.
The reconstruction among them is possible 
with an energy gain of dimerization. \cite{CHADI-CHELIKOW}
The actual process of successive bond breakings and reconstructions
on (100) surface
can be obtained as an unstable cleavage mode. 
\cite{HOSHI2003A}
With the 2D-like constraint, however,
the dimerization is prohibited.
This is why this type of step does {\it not} appear 
within the 2D-like constraint.

The geometries in Fig.~\ref{FIG-STEP-111-2}
can be compared with those proposed in STM experiments.
\cite{FEENSTRA87}
The geometry of Fig.~\ref{FIG-STEP-111-2}(b)
is identical to that of Fig. 5(b) in Ref.\cite{FEENSTRA87},
because of the same alignment of the ring structures;
$5 \rightarrow 7 \rightarrow 6 \rightarrow 5$ before the step
and $6 \rightarrow 5 \rightarrow 7$ after the step.
In the same manner,
the geometry of Fig.~\ref{FIG-STEP-111-2}(c) 
is identical to 
that of Fig. 5(a) in Ref.~\cite{FEENSTRA87}
only before the step ($5 \rightarrow 7 \rightarrow 5$).
After the step, however, the present geometry
($6 \rightarrow 5 \rightarrow 7$)
is different, by one six-membered ring, 
from that shown in Ref.~\cite{FEENSTRA87}
($6 \rightarrow 6  \rightarrow 5 \rightarrow 7$).

The present investigation 
does {\it not} show that
the upper- or lower-type of step in Fig.~\ref{FIG-GEOMETRY}(d) 
is more favorable than the other,
because the present simulations
were carried out under the periodic boundary conditions 
in the $[01\bar{1}]$ direction.
We speculate that 
the appearance of the upper path, 
the \lq$[\bar{2}11]$-type' step, should be enhanced, 
because the step can be formed within 
the minimal crystalline periodicity in the $[01\bar{1}]$ direction.
We have been informed that the two types of step are observed
in STM images \cite{MERA-PRIVATE-COMM}
and we consider that
a more systematic investigation
is desirable both theoretically and experimentally.

\begin{figure*}[thb]
\begin{center}
 \includegraphics[width=14cm]{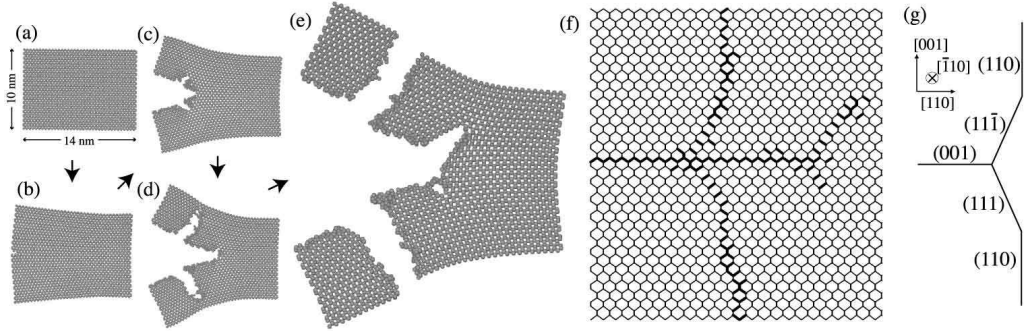}
\end{center}
\caption{
Cleavage process of silicon with bending of cleavage planes.
(a)-(e): Snapshots with a time interval of approximately 1 ps.
Atoms are drawn as balls from the viewpoint of the $[\bar{1}10]$ direction.
(f) : The broken bond sites in snapshot (e) are plotted as bold lines 
in the ideal (crystalline) geometry.
(g) Indices of geometry in (f).
}
\label{FIG-FRAC-BEND}
\end{figure*}

\subsection{Step formation and stability of 
cleavage mode \label{DISCUSSION-CLEAVE}}

The present study of step formation 
can lead to the stability mechanisms 
of the $(111)$-$(2 \times 1)$ cleavage mode,
since a stable cleavage mode 
should be {\it robust} against possible disorders and fluctuations.
We have discussed that 
a step formation is decomposed 
into successive bendings of cleavage plane,
such as $(111) \rightarrow (100) \rightarrow (111)$ planes.
The former and latter bending processes correspond to
the deviation and (quick) recovery
of the $(111)$ cleavage mode, respectively.
Since the $\pi$-bonded $(2 \times 1)$ structure has 
a $\pi$-bonded zig-zag chain 
perpendicular to the cleavage propagation direction,
it can accompany the ordering in the surface structure.
Actually, in Fig.~\ref{FIG-STEP-111-2},
the step formation changes 
the cleaved surface from a disordered (defective) structure
into an ordered $(2 \times 1)$ structure.

Another stability mechanism can be found
in the possibility of multiple propagation directions.
When the $(2 \times 1)$ structures
on the $(111)$ and $(001)$ surfaces are compared,
a crucial difference comes from their symmetry. 
In the limited growth of the $(001)$ cleavage mode,
\cite{HOSHI2003A}
the cleavage propagates easily
in the dimerization direction.
The axis of the dimerization direction is 
unique on the $(001)$ surface 
and a (single) step formation
changes the axis perpendicularly
($(2 \times 1) \rightarrow (1 \times 2)$).
As a consequence,
a cleavage growth with multiple propagation directions
requires step formation
and the resultant $(001)$ cleaved surfaces
tend to be {\it rough} with many steps. \cite{HOSHI2003A}
The steps on the $(111)$ surface, 
on the other hand,
can preserve the cleavage propagation direction
or the direction of the $\pi$-bonded chain
($(2 \times 1) \rightarrow (2 \times 1)$),
as shown in this section.
Moreover, 
the ideal $(111)$ surface, 
unlike the $(001)$ surface,
has three equivalent axes of stable cleavage propagation;
the $[2\bar{1}\bar{1}]$, $[\bar{1}2\bar{1}]$ 
and $[\bar{1}\bar{1}2]$ axes.
In conclusion,
the stability of the $(111)$ cleavage mode 
is supported by the geometrical features;
(i) A step can be formed
without changing the cleavage propagation direction.
(ii) The cleavage propagation direction can be changed 
without step formation. 
The above feature (ii) should result in multiple domains 
without step formation.
Actually, 
an experimental STM image,
Fig. 2 on Ref.~\cite{MERA92},
shows a cleaved $(111)$ surface with multiple domains.
We speculate that 
such a multiple domain structure is formed,
because the cleavage propagation directions are locally different. 

We should emphasize that 
the present stability mechanisms, 
the robustness against step formation
and the possible growth 
with multiple propagation directions,
cannot be explained 
by traditional discussion on surface energy.

\begin{figure}[htb]
\begin{center}
 \includegraphics[width=7cm]{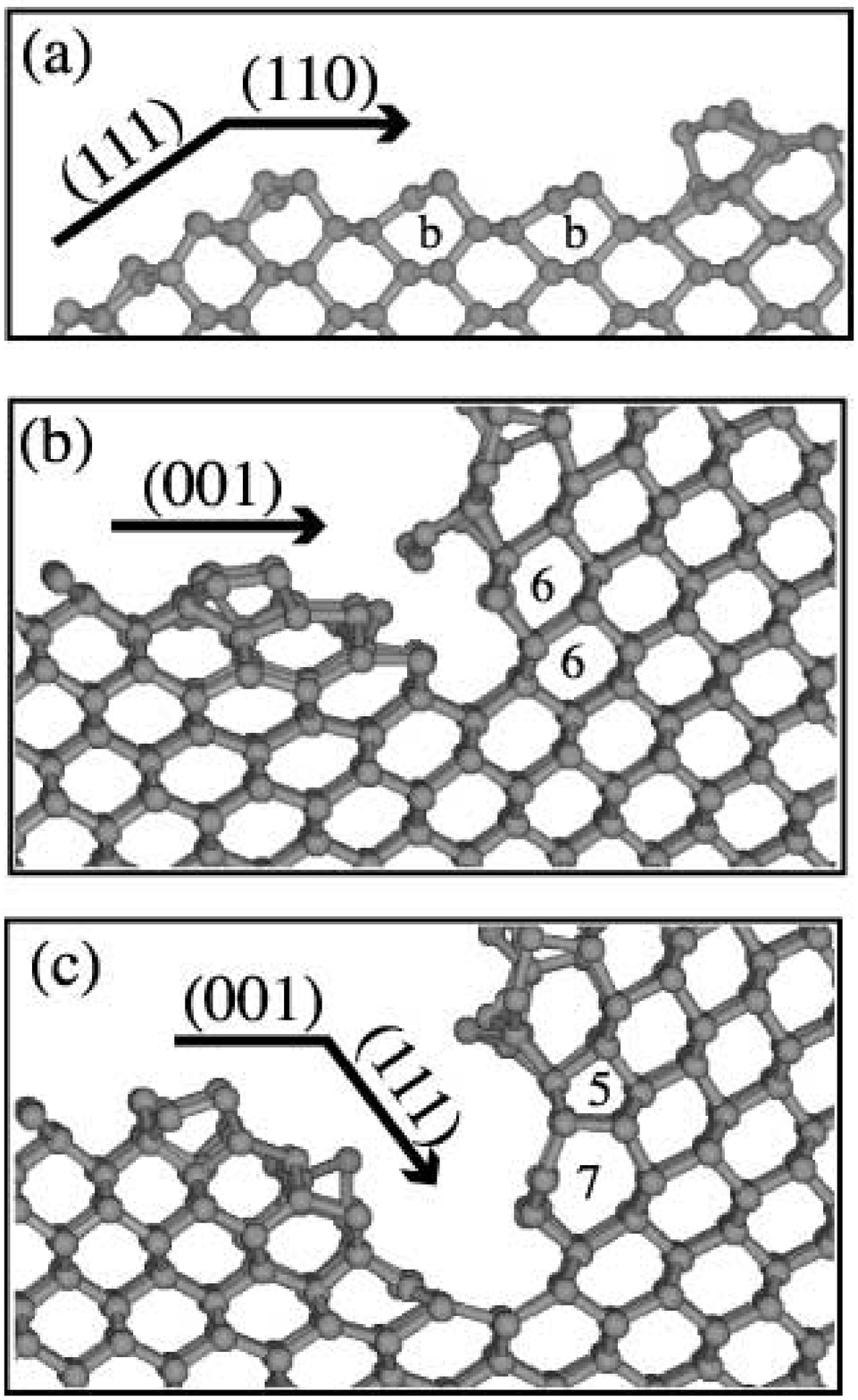}
\end{center}
\caption{
Appearance of well-defined reconstructed surfaces
after bending in cleavage path.
The figures are drawn by projection from the $[\bar{1}10]$ direction;
(a): Bending from $(111)$ plane to $(110)$ plane.
After the bending, the buckled $(110)$ surface structures appears, 
as indicated by \lq b'.
(b)-(c): Two snapshots of the bending 
from $(001)$ plane to $(111)$ plane.
The resultant $(111)$ plane contains 
the $\pi$-bonded $(2 \times 1)$ structure 
with seven- and five- membered rings, indicated as \lq 7' and \lq 5' in (c).
}
\label{FIG-BEND}
\end{figure}

\section{Bending in cleavage path  \label{SECTION-BEND}}

Finally, a direct investigation was carried out
for the preference of 
the $(111)$ or $(110)$ surface in the cleavage process.
We expect that,
even if the simulation initiates the cleavage on other planes, 
the cleavage path will be bent into a more favorable plane.
The above was confirmed by simulations.
Hereafter, symmetrically equivalent planes are 
called by their \lq type'. 
For example, $(1\bar{1}\bar{1})$ and 
$(11\bar{1})$ planes are $(111)$-type planes.

An example
is shown in Figs.~\ref{FIG-FRAC-BEND}(a)-(e). 
The sample contains 10368 atoms 
and the sample size is 
$10 {\rm nm} \times 14 {\rm nm}$ on the $(\bar{1}10)$ plane.
The periodic boundary is imposed 
in the $[\bar{1}10]$ direction by eight atomic layers. 
We prepare an initial \lq seed' of cleavage on the $(001)$ plane.
An external force acts on the atoms near the sample surfaces
except the left one.
See Appendix \ref{DETAIL-CLEAVE} for details.

Figure \ref{FIG-FRAC-BEND}(f) shows 
the broken-bond sites on the crystalline geometry
and the corresponding indices are drawn in Fig.~\ref{FIG-FRAC-BEND}(g).
In this case, 
the bendings are observed from the initial $(001)$ plane 
to $(111)$-type and/or $(110)$-type planes.
The actually observed path shows 
the bending of $(001) \rightarrow (11\bar{1}) \rightarrow (110)$ planes.
Here,
the $(110)$-type planes appear only near the sample surface,
which implies that the appearance of $(110)$-type planes
is enhanced by the present sample geometry.
Different paths were observed 
under different simulation conditions,
such as tuning the magnitude of the external load,
preparing different sample geometry,
changing the region of imposing external force and
setting a different cleavage seed.
For example, 
we observed a bending between two $(111)$-type planes,
such as $(111) \rightarrow (\bar{1}11)$ planes.
In conclusion,
a $(111)$-type or $(110)$-type plane appears,
while no $(001)$-type plane appears except the initially prepared one. 
This conclusion is consistent with 
the experimental preference of 
the $(111)$- and $(110)$-type cleaved surfaces.

Moreover,
the bending path changes, sometimes,
from a disordered (defective) area 
into an ordered one
with a well-defined reconstruction;
In Fig.~\ref{FIG-BEND}(a), 
the $(110)$ plane after bending
contains the buckled structures,
known for typical reconstruction. \cite{STEKOL-2}
In the process shown in Figs.~\ref{FIG-BEND}(b)-(c),
the resultant $(111)$ cleavage plane 
contains the $\pi$-bonded $(111)$-$(2 \times 1)$ structure.



\section{General discussion \label{DISCUSSIONS}}

The present investigation of cleavage 
can be discussed from 
the general viewpoints of (I) time scale and (II) length scale.
From the viewpoint of time scale (I),
the metastable Si$(111)$-$(2 \times 1)$ structure appears,
owing to the local reconstruction mechanism 
within a limited time scale.
Since the mechanism is general 
in the quantum mechanical picture,
we expect the $(111)$-$(2 \times 1)$ surface also 
in other nonequilibrium processes.
Actually, 
we were informed that 
the corresponding STM image 
is found in a Si(111)$\sqrt{3}\times\sqrt{3}$-Ga surface,
when the Ga atoms are locally removed 
at room temperature. \cite{NOTE-ICHIKAWA}
At higher temperatures,
the $(7 \times 7)$ structure appears. \cite{ICHIKAWA-SURF}
The above situation 
is another \lq fast' process almost free from thermal equilibration.
A review article \cite{NEDDER} lists
other experiments with the appearance of the Si(111)-$(2 \times 1)$ structure.

From the viewpoint of length scale (II), 
structure of material is generally determined 
by volume and surface terms of energy.
Dimensional analysis shows a crossover of mechanical property 
between nano- and macroscale systems;
The two energy terms are competitive in nanoscale system,
while the volume term is dominant in macroscale system.
In the case of silicon nanostructure,
the above energy competition originates from 
that between the $sp^3$ (bulk) term
and the non-$sp^3$ (surface) term.
In the present investigations,
the crossover at the 10 nm scale is found 
in step formation or bending in cleavage path.
We speculate that 
the crossover at the 10 nm scale is universal in silicon, 
because the surface energy
is always on the same order among different surface 
indices and reconstructions.\cite{NOTE-SURFACE-ENERGY}
The crossover may be seen also in 
the size dependence of  
the shape in self-organized Si islands; \cite{ICHIKAWA2002}
Islands with a size of 10 nm or less
have a semispherical shape and 
those with a larger size have 
a pyramidal shape with facets in well-defined indices.
These crossovers at the 10 nm scale can be understood,
because a well-defined reconstructed surface appears,
only when the system size is much larger than 
the unit length of the reconstructed surface ($\sim$1 nm).

The present method of large-scale electronic structure calculation
has wide applications and is not specific to cleavage.
The structural property at the 10 nm scale 
is an urgent problem of the present technology 
and can be a future study of the present method.

\appendix

\section{Methods for large-scale electronic structure calculation 
\label{APPENDIX-A}}

\begin{figure}[thb]
\begin{center}
 \includegraphics[width=7cm]{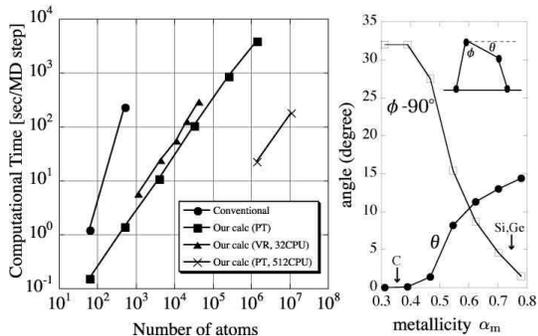}
\end{center}
\caption{
(Left) Computational time 
for molecular dynamics simulation with up to 11,315,021 Si atoms.
Our calculations are compared with the conventional calculation
for the eigenstates.
See text for explanations.
(Right) Dimer geometry on (001) surface 
of group IV elements (C, Si and  Ge).
The angle $\theta$ is the tilt angle and
$\phi$ is the angle
between the surface dimer and 
the plane that contains the two back bonds of the upper atom.
}
\label{FIG-BENCHMARK}
\end{figure}

In recent years,
we have developed theories and program codes
for large-scale electronic structure calculations.
\cite{HOSHI2000A,HOSHI2001A,HOSHI2003A,GESHI,TAKAYAMA2004A,THESIS}
Among them, 
a variational (VR) procedure 
for generalized Wannier states \cite{HOSHI2000A,THESIS}
is used in the present cleavage simulations.
The Wannier state is well-defined 
localized \lq chemical' wavefunction in condensed matter, 
such as a bonding state or a lone-pair state, 
with a slight spatial extension.
The suffix $i$ of a Wannier state $\phi_i$
denotes its localization center.
A bench mark is shown in 
the left panel of Fig.~\ref{FIG-BENCHMARK},
in which our calculations are compared with 
the conventional method for calculating eigenstates.
The figure shows not only the result of the VR procedure but also 
that of another Wannier state method called  
\lq perturbative (PT) procedure'. \cite{HOSHI2000A,HOSHI2003A}
The circle and square indicate the results of 
the conventional method and PT procedure, respectively, 
by a standard workstation. \cite{HOSHI2003A} 
The triangle and cross indicate the results 
of the VR procedure with 32 CPUs 
and PT procedure with 512 CPUs, respectively, 
by a parallel computation system 
(SGI Origin 3800$^{\rm TM}$).
The parallelism is carried out
by the OpenMP technique (www.openmp.org).
All the results of our methods are
\lq order-$N$', or 
linearly proportional to the system size ($N$). 
The PT procedure is much faster than the VR procedure
but its applicability  is strictly limited.
\cite{HOSHI2000A,HOSHI2003A}
Several related methods \cite{HOSHI-HYBRID} were also developed 
but are not used in this paper.

Another methodological foundation
is the transferable Hamiltonian \cite{KWON}  
in the Slater-Koster (tight-binding) form,
applicable to various circumstances,
{\it e.g.}, crystals, liquid, defects and surfaces.
Its success is based on the universality of electronic structures, 
which has been known for decades \cite{PHILLIPS,HARRISON,VOGL}
and can be founded by the {\it ab initio} theory. 
\cite{LMTO-HIGHLIGHT,NMTO,NMTO-2}
Consequently,
group IV elements can be systematically described 
by a one-parameter energy scaling theory.
\cite{PHILLIPS,HARRISON,HOSHI2000A,HOSHI2001A}
The scaling parameter, 
\lq metallicity' $\alpha_{\rm m}$, is defined as
\begin{eqnarray}
\alpha_{\rm m} \equiv \frac{\varepsilon_{p} - \varepsilon_{s}}{2t_{sp^3}}.
 \label{METALLICITY}
\end{eqnarray}
Here, 
$(\varepsilon_{p} - \varepsilon_{s})$ is the energy difference 
between the $p$ and $s$ orbitals and
$t_{sp^3}$
is the transfer energy along a bulk ($sp^3$) bond
$(t_{sp^3} \equiv |V_{ss \sigma}-2\sqrt{3} 
V_{sp \sigma} - 3 V_{pp \sigma} |/4)$.
A typical value for C is
$\alpha_{\rm m} =0.35$.\cite{XU} 
The values for Si and Ge are similar,
$\alpha_{\rm m} =0.75-0.78$
(see Refs.~\cite{HOSHI2000A,HOSHI2001A}
and references therein).

Several surface structures among C, Si and Ge
are systematically explained by the energy scaling theory.
An example is shown in 
the dimer geometry on (001) surfaces.
{\it Ab initio} calculations 
result in a {\it symmetric} dimer in the C case 
and similar {\it asymmetric} dimers
in the Si and Ge cases
(see Fig. 2 of Ref.~\cite{KRUGER95}, for example).
In the right panel of Fig.~\ref{FIG-BENCHMARK},
the above trend is reproduced with the present Hamiltonian  
by tuning the value of $\alpha_{\rm m}$
or $(\varepsilon_{p} - \varepsilon_{s})$.
Since the $(111)$-$(2 \times 1)$ cleaved surface
is commonly seen in Si and Ge\cite{NOTE-GE-111},
the phenomena should hold a common mechanism
in the present context of universality.

For validation of the present study,
the surface energy of several Si surfaces was 
calculated by the present Hamiltonian as follows
and the {\it ab initio} values
\cite{STEKOL-2} are given inside the parentheses;
$\gamma_{001}^{4 \times 2}= 1.58 \, (1.41)$ J/m$^2$ ,
$\gamma_{111}^{2 \times 1}= 1.97 \, (1.44) $ J/m$^2$ and
$\gamma_{110}^{\rm buckled}= 2.11 \, (1.70)$ J/m$^2$,
for 
the $(001)$-c$(4 \times 2)$ structure,
the $\pi$-bonded $(111)$-$(2 \times 1)$ structure and
the buckled $(110)$ structure, respectively.
The present Hamiltonian reproduces
the {\it ab initio} results satisfactorily, 
particularly, the order of magnitude
($\gamma_{001}^{4 \times 2} < 
\gamma_{111}^{2 \times 1} <
\gamma_{110}^{\rm buckled}$),
though the absolute values are somewhat overestimated.
The calculated surface energy of $(001)$ surfaces also gives
satisfactory results. \cite{FU}

%
%

%
\section{Conditions of cleavage simulation \label{DETAIL-CLEAVE}}

This appendix explains the technical conditions
for cleavage simulation.
The time step of the molecular dynamics was set at 3 fs. 
The center of gravity of the sample was fixed as a constraint.
Sample surfaces 
were terminated by orientationally fixed $sp^3$ bonding states.
This boundary condition corresponds to the situation in which 
the sample is embedded in a bulk ($sp^3$-bonded) system.  
This situation is quite different from
that with usual hydrogen termination, 
in which the atomic structure is deformed significantly 
from the tetrahedral geometry, 
owing to the large deviation from the $sp^3$ bonding. 

Initial defect bonds were prepared for the cleavage \lq seed',
which can be seen in Fig.~\ref{FIG-111-LARGE}(b) or Fig.~\ref{FIG-FRAC-BEND}(b). 
Actually,
an additional short-range repulsive force was imposed
on the atom pairs of selected bond sites.
The selected bond sites were,
typically, two or three bond sites.
Since the repulsive force is of short range,
it will act on nothing after the bond breaking.

Here,
we discuss the way of imposing the external load
on sample surfaces (boundaries).
If the sample is sufficiently large,
the cleavage phenomena will be determined 
by the intrinsic nature of the solid,
not by the boundary condition of the sample.
We have investigated various ways of 
imposing an external load \cite{THESIS}
and the discussion in this paper is based on the results that 
are {\it not} sensitive to the boundary conditions.
We pick out the cases of 
Figs.~\ref{FIG-111-LARGE} and \ref{FIG-FRAC-BEND}.
In Fig.~\ref{FIG-111-LARGE}, 
the external load is realized
by the constrained movement.
The atoms of the outermost two layers
on the top and bottom sample surfaces
were constrained under artificial movement
in the $[111]$ direction.
Since the cleavage propagation velocity
is on the order of the Rayleigh wave velocity $v_{\rm R}=4.5$nm/ps,
the velocity of the artificial movement should be much slower 
than $v_{\rm R}$.
If not, the atoms in the constraint motion 
(the atoms near the sample surfaces) 
are removed from the sample
and no internal (cleavage) surface appears.
In the case of Fig.~\ref{FIG-111-LARGE},
we chose $v_0=0.16{\rm nm/ps}$, as a typical value.
In our simulations, 
the artificial movement was required
to avoid forming multiple cleavage planes.
The details are described elsewhere. \cite{THESIS}
Within the above requirements for the artificial movement,
the conclusions of the present paper are not changed.
The way of controlling the artificial movement 
is reflected on the shape of the sample boundary 
in Fig.~\ref{FIG-111-LARGE} (a)-(e).

Now we explain 
the case of Fig.~\ref{FIG-FRAC-BEND}, in which 
an external force field is imposed 
on selected atomic layers near the sample surfaces, except the left one.
The thickness of the selected layers is 5 or 10 \% of the sample length. 
The force field is given as 
\begin{eqnarray}
 \bm{F}(\bm{r}) = \frac{K d_0^2}{\sqrt{2 \pi r}} \frac{\bm{r}}{|\bm{r}|}
 \label{EXT-FORCE}
\end{eqnarray}
with the relative  coordinate $\bm{r}$ from the force center.
The (two-dimensional) coordinate $\bm{r}$ is given 
on the $(01\bar{1})$ plane (See Fig.~\ref{FIG-GEOMETRY}).
The force center ($\bm{r}=0$) is fixed at 
the position of the initial defect and is placed almost 
at the middle of the left sample surface.  
The length $d_0$ 
is determined so that the volume per atom is given as
$d_0^3$ ($d_0 \approx 3$\AA).
The force field is consistent with
the concentrated stress field 
given by the continuum mechanics of cleavage, \cite{BRITTLE}
when its angular dependence is ignored.
The angular dependence was ignored in our simulations,
because it is defined only for a planar crack 
and is not valid in the stepped or bent cleavage paths discussed in this paper.
In Fig.~\ref{FIG-FRAC-BEND}, 
$K = 0.8 {\rm MPa \sqrt{m}}$,
which is slightly larger than 
the critical value ($K_{\rm prop} = 0.7 {\rm MPa \sqrt{m}}$)
estimated in Appendix \ref{TOUGHNESS}.
In several other simulations,
the field center was under artificial movement,
owing to the fact that 
the crack tip moves with the cleavage propagation. 
We did not find, however, a systematic difference 
in the resultant cleavage behavior.

Finally, 
in Figs.~\ref{FIG-111-LARGE} and \ref{FIG-FRAC-BEND}, 
the total kinetic energy 
was controlled by the thermostat method \cite{NOSE}
with a typical temperature parameter of $T$=800K.
The temperature parameter $T$, however,  
does {\it not} correspond to experimental temperature, 
since the simulation is a nonequilibrium process.
In Fig.~\ref{FIG-111-LARGE}, for example,
the constraint movement is introduced to the sample surfaces
with the characteristic velocity $v_0$.
The temperature parameter $T$ should be sufficiently large
to propagate the introduced deformation into the internal region.

\section{Estimation of critical stress intensity factor \label{TOUGHNESS}}

The critical stress intensity factor for cleavage propagation,
$K_{\rm prop}$, 
was estimated for the $(111)$-$(2 \times 1)$ cleavage mode 
with a smaller sample of $416$ atoms.
In the continuum theory with linear elasticity,\cite{BRITTLE}
the divergent stress field is presented at the crack tip.
The center of the divergent field $\bm{R}_{\rm c}$ is given and 
the stress intensity factor $K$ is introduced
as the amplitude of the divergent field.
Its critical value $K_{\rm prop}$ is 
a measurement of the critical external load 
and was determined in several experimental techniques and 
in the electronic structure calculations 
with hundreds of atoms. ~\cite{SPENCE93,GUMBSCH}

The present estimation of the critical factor $K_{\rm prop}$
was carried out within the following quasi-static picture;
when the field center $\bm{R}_{\rm c}$ is shifted 
by one atomic increment,
the crack tip should be shifted, also by one atomic increment,
with the elementary bond-breaking and reconstruction process.
If the factor $K$ is not sufficiently large $(K < K_{\rm prop})$, however,
the crack tip is not shifted.
In practical simulations,
the field center $\bm{R}_{\rm c}$ was chosen 
in the middle of a bond layer,
which is shown as the nonstepped path of 
Fig.~\ref{FIG-GEOMETRY}(d).
As constraints, 
outer atoms were fixed with 
the linear elastic displacement of plane strain~\cite{BRITTLE,SPENCE93},
determined by the factor $K$ and the position $\bm{R}_{\rm c}$.
Internal (movable) atoms are relaxed 
in the finite temperature of 300 K for 2.4 ps.
After the relaxation of the internal atoms,
the position $\bm{R}_{\rm c}$ is shifted 
by $1/10$ of one atomic increment and 
the relaxation is restarted.
The resultant cleaved surface shows the buckled $(2 \times 1)$ structure,
as in Fig.~2 of Ref.~\cite{SPENCE93}.
In our simulations,
the increase of the reconstructed surface area by one atomic increment
occurs within the typical time scale of 0.1 ps.

The present calculations resulted in 
a typical value of $K_{\rm prop} = 0.7 {\rm MPa \sqrt{m}}$,
which agrees satisfactorily 
with the previous theoretical  and experimental values of
$0.41 - 1.24 {\rm MPa \sqrt{m}}$ listed in Table I of Ref.~\cite{SPENCE93}
Since the calculated surface energy 
in Appendix \ref{APPENDIX-A}
is somewhat overestimated,
for example, by approximately 40 \% in $\gamma_{111}^{2 \times 1}$,
we speculate that the present value of $K_{\rm prop}$
is an overestimated one,
owing to the limited freedom of the present Hamiltonian.
\cite{NOTE-GRIFFITH}

\begin{acknowledgements}

We thank Yutaka Mera, Kouji Maeda and Masakazu Ichikawa 
(University of Tokyo) for discussion on experiments.
Numerical calculation was partly carried out 
using the facilities of the Japan Atomic Energy Research Institute
and the Institute for Solid State Physics, University of Tokyo. 
This work is financially supported 
by \lq Research and Development for Applying advanced 
Computational Science and Technology' 
of the Japan Science and Technology Corporation. 
\end{acknowledgements}



\begin{thebibliography}{99}
\bibitem{NOTE-SURFACE-ENERGY}
The surface energy of silicon
is always given on the order of 
1 J/m$^2$ or 1 eV per atom.
\bibitem{NOTE-GRIFFITH}
In the continuum theory, 
the critical value of the stress intensity factor $K$ 
is proportional to 
the square root of surface energy
($K \propto \sqrt{\gamma}$).
\cite{GRIFFITH,BRITTLE,SPENCE93}
\end{thebibliography}
\end{document}